\documentclass[preprint,aps]{revtex4-1}
\usepackage{wrapfig}
\usepackage{graphicx}% Include figure files
\usepackage{dcolumn}% Align table columns on decimal point
\usepackage{bm}% bold math
\usepackage{hyperref}
\usepackage{epsf}
\usepackage{float} 
\newcommand{\eq}[1]{(\ref{#1})}
\newcommand{\la}{\label}
\newcommand{\eea}{\end{eqnarray}}
\newcommand{\beq}{\begin{equation}}
\newcommand{\eeq}{\end{equation}}
\newcommand{\be}{\begin{equation}}
\newcommand{\ee}{\end{equation}}
\newcommand{\ii}{{\rm{i}}}

\newcommand{\vv}{{\rm v}}
\newcommand{\uu}{{\rm u}}
\newcommand{\pp}{{\rm p}}
\newcommand{\p}{\partial}

\def\XXint#1#2#3{{\setbox0=\hbox{$#1{#2#3}{\int}$ }
\vcenter{\hbox{$#2#3$ }}\kern-.5\wd0}}

\usepackage{amsmath}
\usepackage{amsfonts}
\usepackage{varioref}

\begin{document}

\title{Anomalous  Hydrodynamics of Fractional Quantum Hall States\\ }% Force line breaks with \\

\author{P. Wiegmann }
 \affiliation{ Department of Physics, University of Chicago, 929 57th St, Chicago, IL 60637}%Lines 
\date{\today}
%  automatically or can be forced with \\
%  \email{Second.Author@institution.edu}
% \affiliation{%
%  \homepage{http://www.Second.institution.edu/~Charlie.Author}
% \affiliation{

\date{\today}% It is always \today, today,

\begin{abstract}
We propose a comprehensive framework  for the quantum  hydrodynamics of  the Fractional Quantum
Hall (FQH) states. We suggest that the electronic fluid in the FQH regime could be phenomenologically  described by the quantized hydrodynamics of vortices in    an incompressible rotating liquid. We     demonstrate  that such hydrodynamics   captures all major features  of FQH states including  the subtle  effect of  Lorentz shear stress. We present a  consistent quantization of  hydrodynamics of an incompressible fluid providing a powerful framework to study FQHE and superfluid. We obtain the quantum hydrodynamics
of the vortex flow by quantizing the  Kirchhoff equations for vortex dynamics. \end{abstract}

\maketitle
\section{ Introduction}    Quantum systems with effectively strong interaction form  liquids whose flows are  coherent quantum collective motions. Among them there are interesting notable cases when such liquids allow a \emph{hydrodynamics description}. That is when the long wave, slow flows can be effectively  described solely in terms of a  macroscopic, but quantum, pair of canonical fields of density \(\rho(r,t)\) and velocity \(v(r,t)\).
Such quantum flows are the subject of quantum hydrodynamics. In  the classical case  the  principle  of local equilibrium  reduces the Boltzmann kinetic equation for the distribution function to the hydrodynamics equations for the density and the velocity (see, e.g.,\cite{LandauX}). Local equilibrium occurs when the characteristic  time  of the flow exceeds  the characteristic time of collisions, and  the characteristic scale of the flow exceeds the  mean free path of particles. A quantum  analog of the principle of local equilibrium, is yet to be  understood, but when it comes to effect,    involves a long range coherent effects. Strong coherence  emerges as a result of   interactions. Noticeable examples of quantum hydrodynamics are superfluid helium, superconductors, trapped cooled atomic gases, and Luttinger liquids.  A Fractional Quantum Hall (FQH) liquid is yet another case.

 Electronic states   confined within the lowest Landau
level by  quantizing magnetic field are holomorphic. The   holomorphic nature of states make the hydrodynamic description possible. 

A quest for the hydrodynamics of a FQH liquid    originated
in a seminal
paper \cite{GMP}. Earlier approaches to FQH states  of  Refs \cite{Kivelson,Dung,Read89}
are somewhat related to the hydrodynamics, as  noted in Ref. \cite{Stone}.
Hydrodynamics of FQH states is in the  focus of  a renewed  interest.

 In hydrodynamics,   a few basic principles, symmetries,  and
a few phenomenological parameters are sufficient to formulate the   fundamental
equations. 
In the case of FQHE we already possess sufficient characterizations of states.  They can be used 
as a basis of the hydrodynamics approach.
To this aim a microscopical Hamiltonian and a deeper understanding
of  underlying  microscopic  mechanisms of emergence of correlated liquid
states are, in fact,  not necessary.

In this paper we  formulate a minimal number of   principles sufficient  to
develop  the hydrodynamics
of  FQH bulk states in a  close analog of Feynman's theory of superfluid
helium \cite{Feynman}, and the  theory of collective excitations
 in FQH states of Ref. \cite{GMP}.
We discuss only the simplest
Laughlin states.
Elsewhere, we hope to be able to address the  hydrodynamics of other, richer  FQH states,  possessing additional symmetries,
such as the $5/2$ state. They can be studied     within the framework developed here.

We  argue that states of the  FQH liquid can be treated as \begin{center}\emph{Flows
of quantized vortices in a quantum
incompressible rotating inviscid  liquid}.\end{center}
On this basis we obtain
the major  features of FQHE including subtle effects such as  Lorentz shear stress
\cite{HV},  missed by the previous
 approaches \cite{Kivelson,Dung,Read89,Stone}.
In particular, the Laughlin wave function
\begin{align}
 \psi_0(z_1,\dots,z_N)=e^{-\frac{1}{4\ell^2}\sum_i|z_i|^2}\prod_{i>j}(z_i-z_j)^\beta
,\quad\quad\beta=\frac 1\nu.\end{align}  
emerges as the ground state of the vortex fluid.
Here \(\ell=\sqrt{\hbar
c/e B}\) is the magnetic length. 

To  the author's knowledge,  a hydrodynamics of   vortex flows  has not been developed. It is an interesting
subject of its own. Apart from FQHE it is also relevant to the theory of
superfluids and classical hydrodynamics. In this paper we present a consistent quantum hydrodynamics of such a fluid.

Hydrodynamics of the vortex fluid differs from the Euler hydrodynamics. Its quantum version differs from the canonical quantum  hydrodynamics of Landau
\cite{Landau}. The major difference is 
 the {\it anomalous terms}. These terms  represent the Lorentz shear
force. Emergence of such forces in hydrodynamics, classical and quantum alike, is the
major focus of this paper. 

We start from the observation that the FQH states can be interpreted as the states of quantized Kirchhoff vortex matter, Sec.\ref{K}, and then develop the hydrodynamics of the vortex matter in Sec \ref{V}.
We summarize the main results   in the Sec. \ref{main}, followed by the  details of derivations presented in Sec.\ref{R}-\ref{X}.

 Some  results presented below were
obtained in collaboration with Alexander Abanov.
This paper is an extended version of Ref. \cite{W2} written for the special issue of JETP dedicated to Anatoly Larkin.
Discussions of hydrodynamics of quantum liquids with 
I. Rushkin, E. Bettelheim
and T. Can  and their help in understanding material presented below are acknowledged.

The  author
 thanks the Simons Center for Geometry and Physics   for the hospitality during the
completion of the paper. The work was supported by NSF DMS-1206648, DMR-0820054
and BSF-2010345.

\section{foundational principles of
hydrodynamics of FQH  liquid}\la{F}
\subsection{Characterization of Fractional Quantum Hall states }

Electrons in a quantizing magnetic field confined in 2D heterostructures in the regime dominated by the Coulomb interaction form  FQH states. The most robust  FQH states occur at the filling fraction ? = 1/3, that are the Laughlin states. The FQH states form a quantum liquid. Characterization of that liquid is:
\begin{itemize}
\item [-] 
 Flows are incompressible \cite{L}, and almost  dissipation-free \cite{T,T1};
\item [-] The spectrum of bulk excitations  is gapped 
\cite{T,GMP}.    The gap is less than the cyclotron energy \(\hbar\omega_c>\Delta_\nu\). Only edge states,  excitations  localized on the boundary, are soft \cite {edge};

\item [-] The Hall conductance  is  fractionally quantized 
\cite{T1};

\item [-]  Elementary excitations  in the bulk  of the fluid are vortices. Vortices carry  fractionally quantized negative electronic charge \cite{L}.
\end{itemize} More subtle features recently discussed in the literature  are:

\begin{itemize}
\item 
[-]Edge excitations consist of two branches of  non-linear
solitons: subsonic solitons
with the  fractional negative electronic charge and supersonic solitons with the unit electronic charge \cite{W};
\item [- ]Quantized   double layers of the density  at  boundaries and vortices \cite{W};
 \item[-]  The Lorentz shear stress and anomalous  viscosity (odd viscosity or Hall viscosity) \cite{HV}. 
\end{itemize}
From the listed properties we select a  set of the   foundational  principles and  attempt to obtain others as consequences. The set of basic principles is remarkably minimal. We  only assume that electrons in the FQH regime form a quantum fluid, and that

\begin{itemize}
\item 
[] \emph{The  fluid is incompressible, and  flows possess a macroscopic number of equally oriented vortices.}
\end{itemize}   

We refer to such flow as a chiral flow. Since in the quantum fluid vorticity is quantized, a unit volume of the fluid contains the quantum of vorticity. We want to demonstrate that the chiral flow captures all known physics of FQHE.

 We start with a general discussion of scales of FQH (bulk) states.

\subsection{Scales, Holomorphic States, Incompressibility}  There  are two distinct energy scales in the physics of FQHE: the cyclotron energy \(\hbar\omega_c = e\hbar B / (m_bc) \),  which provides the  distance between Landau levels, and  the gap in the bulk excitation spectrum \(\Delta_\nu\).  The former is determined by the band electronic mass \(m_b\) and by the  magnetic field. The  latter is the characteristic of the  Coulomb energy. From the theoretical standpoint, the very existence of FQH states assumes that the cyclotron energy is  larger than the gap \(\Delta_\nu\ll
\hbar\omega_c\). If this limit holds \cite{mass}, the flows with   energy   \(E\)    exceeding  the gap may  still be comprised of  states on the lowest Landau level \(\Delta_\nu\ll
E\ll\hbar\omega_c\).  Such motion does
not depend on the band  electronic mass \(m_b\).

Let us consider a small modulation of electronic density \(\rho(r)\) and ignore for the moment the
electrostatic interaction of a nonuniform charged fluid \cite{electrostatics}.  Such   flow
 has a momentum flux \(P(r)\) and propagates with velocity \(v(r)\).  
We assume that  at small modulations  the momentum flux  is equal  to  \(P=m_*\rho v\), where \(m_*\) is the inertia of the flow.   It seems  naturally to assume that the inertia is set by the scale  provided by the gap      \(\Delta_\nu\sim\hbar^2/(m_*\ell^2)\). The mass \(m_* \) exceeds that of a band electron \(m_*/m_b\sim (\hbar\omega_c)/\Delta_\nu >1\).

 Waves propagating through the bulk  of the FQH liquid are essentially non-linear.  Linear waves in the bulk  are possible as a response of the ground state to    electric and magnetic fields and to spatial curvature.  That sector of states  often called {\it topological}.
 
 In the paper we consider a setting when the electronic droplet occupies a finite volume confined by a smooth potential well.
The wave functions of states with energy  less than  the cyclotron energy (the lowest Landau level) are holomorphic.  It is customary  to describe  the set of  states on the lowest Landau level as the Bargmann space \cite{Bargmann}.   Coherent states of the Bargmann space are labeled by symmetric  polynomials of the holomorphic coordinates of particles  \(z_i=x_i+\ii y_i\) and the holomorphic momenta \(\p_{z_i}=\frac{1}{2}(\p_{x_i}-\ii\p_{y_i})\).  Let \(Q\) be such a polynomial, and \(Q^\dag\)  the Hermitian  conjugated polynomial which depends on anti-holomorphic coordinates \(\bar z_i=x_i-\ii y_i\), and anti-holomorphic momenta
\(\p_{ z_i}^\mathrm{\dag}=-\frac{1}{2}(\p_{x_i}^\mathrm{T}+\ii\p_{y_i}^\mathrm{T})\).  The symbol \(^\mathrm{T}\) is the transposition. Then in the notations
of the Bargmann space the "bra" and "ket" states are\begin{align}\la{Q1}
\langle Q|=\overline {}\prod_{i>j}(\bar
z_i-\bar z_j)^\beta\cdot Q^\dag e^{-\frac{1}{2\ell^2}\sum_i|z_i|^2},\quad Q\cdot\prod_{i>j}(
z_i-z_j)^\beta=|Q\rangle
\end{align}
 Flows  within the first Landau level are incompressible. Sometimes the term "incompressible flow  " is    attributed to the gapped spectrum. Rather, the incompressibility   reflects the holomorphic nature of FQH-states. This is seen from the following argument. For simplicity, let us consider a coherent state characterized by a  polynomial  \(Q\) which depends only on coordinates \(z_i\) . The phase of the wave function of such state  differs from the phase of the ground state by the phase of the holomorphic polynomial   \({\rm  Im}\log Q\).  Since the velocity is a gradient of the phase, the phase is the hydrodynamic potential. The phase is harmonic everywhere except points where the wave function vanishes. Since the wave function is single valued it vanishes as an integer power of holomorphic coordinates. Therefore the  allowed singularities of the phase  correspond to quantized vortices. There are no sources so  the gradient of the phase is  divergence-free\begin{align}\label{i1}\omega_c\to\infty:\quad\quad\qquad
 \nabla\cdot{  v}=0.
\end{align}
There are two immediate
consequences of incompressibility. One is that  the material derivative of the density vanishes
 
\begin{align}\la{E1}D_t\rho\equiv\left(\frac{\partial}{\p t} +
  v\cdot \nabla\right)\rho=0.%&\rho+\mathbf{\nabla( v}\cdot\rho)=0.
\end{align} 
  the second is that that flows
in homogeneous   2D  incompressible
 liquids do not possess  linear  waves. Only available bulk flows
are non-linear     flows of vorticity \cite{LV}.  The flow
can be viewed as a motion of a neutral gas of quasi-holes and quasi-particles. 

In the next section we identify the the FQH states with vortices in a quantum incompressible rotating fluid. 
\section{Kirchhoff Equations} \la{K}
We start by recalling classical Kirchhoff equations for the rotating incompressible
inviscid Euler flows  with a constant density (see e.g.,\cite{Newton}), and  then proceed with the quantization.

\subsection{Classical Kirchhoff Equations for an incompressible fluid}
In two dimensions an   incompressible fluid with the constant density is fully characterized by its
vorticity.  The curl of the Euler equation for the incompressible fluid with a constant density\begin{align}\la{EE}
 D_t u\equiv(\p_t+u\cdot\nabla) u=-\nabla p
\end{align}
yields the single (pseudo) scalar equation for the vorticity
\begin{align}\la{E}{D_t (\nabla}\times
  u)=0.
  %&\rho+\mathbf{\nabla( v}\cdot\rho)=0.
\end{align}In this form the Euler equation  has a
simple geometrical meaning: the material derivative
of the
vorticity vanishes. Vorticity is  transported
 along    divergence-free velocity field \(  u\).

  \begin{wrapfigure}[13]{L}{0.3\textwidth}
\includegraphics[width=0.3\textwidth]{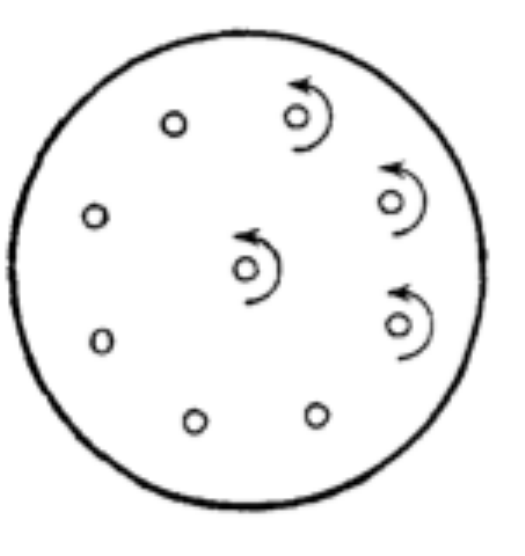}
\caption{ A  picture from the Feynman book \cite{Feynman}
illustrating a chiral flow}
\end{wrapfigure}
  Helmholtz, and  later Kirchhoff realized that there is a class
of solutions of the vorticity equation \eq{E} which consists of a finite
number of  point-like vortices. In this case  the
complex velocity of the fluid  \(\uu=u_x-iu_y\)  is the meromorphic function 
 \begin{align}\label{k}
\uu(z,t)=-\ii \Omega\bar z+\ii\sum^N_{i=1 }\frac{\Gamma_i}{z-z_i(t)} ,
\end{align}
where  \(\Omega\) is an angular velocity  of the rotating  fluid, \(N\)
  is the   number of vortices,  \(\Gamma_i\) and   \(z_i(t) \) are circulations
and positions
of vortices.   

A substitution of this "pole Ansatz" into the Euler equation \eq{E}
yields that the number of vortices \(N\) and the circulations    
\(\Gamma_i\) do not change in time,   while the moving positions
of vortices \(z_i(t)\) obey the  Kirchhoff equations:
\begin{align}\label{32}
  \dot {\overline z}_i=-\ii \Omega\bar z_i+\ii\sum^N_{i\neq j
}\frac{\Gamma_j}{z_i(t)-z_j(t)}.
\end{align}The Kirchhoff equations replace the non-linear PDE \eqref{E} by a dynamical
 system. They can be used for
different purposes. 
Equations describe chaotic motions of a finite number of
vortices if \(N>3\). If $N$ is large  Kirchhoff equations  can be used  to  approximate virtually any flow.
\subsection{Chiral flow}
 The flows  relevant for FQHE  are such that    a large number of  vortices     largely compensates
rotation.
We refer such flows as the  {\it
chiral flow.} 
Bearing in mind
the quantum case we assume that vortices have the same (minimal) circulation
 \(\Gamma_i=\Gamma\).  
Then Kirchhoff equations read\begin{align}\label{u}
\vv_i\equiv  \dot {\overline z}_i=-\ii \Omega\bar z_i+\ii\sum^N_{i\neq j
}\frac{\Gamma}{z_i(t)-z_j(t)}.
\end{align}
 We want to study the vortex system in the limit of large number of vortices
distributed
with the mean density  \(\bar\rho\) 
\begin{align}
    N\to\infty:\quad \quad \bar\rho=\/\frac{\Omega}{\pi \Gamma}. 
\end{align}  The chiral flow is a very special flow in fluid mechanics. There we distinguish
two types of motion: {\it fast motion} of the fluid around vortex cores, and a
{\it slow motion} of vortices.  In this respect vortices  themselves may be considered
as a (secondary) fluid. In the ground state of  the vortex fluid  vortices do not
move, but the fluid does. 
 
   Circulation of vortices  in units of the Plank constant has dimension  inverse  to  the mass unit. We set the dimensionless parameter
\begin{align}\la{321}
\nu=\frac{\hbar}{m_*\Gamma}\;.
\end{align}
 Later we show that the quantized chiral flow  models FQHE with a filling fraction \(\nu\).
We denote \(\beta=\nu^{-1}\).

\subsection{Quantum Kirchhoff Equations}\la{C1} Kirchhoff himself wrote equations \eq{u} \ in the Hamiltonian form
identifying holomorphic and anti-holomorphic  coordinates of vortices as canonical variables.  In the case of the rotating fluid the 
the canonical variables are \(m_*\Omega\bar z_i\), and \(z_i\). The Hamiltonian of the chiral vortex system is\begin{align}\la{36}
 \mathcal{H}=m_*\Omega\left(\sum_i[\Omega|z_i|^2-\Gamma\sum_{j\neq i}\log|z_i-z_j|^2\right),\quad
(m_*\Omega)^{}\{\bar z_i,\,z_j\}_{P.B.}=-\ii\delta_{ij}.
\end{align}  The
parameter \(m_*\) we put in the Hamiltonian and the Poisson brackets sets
the scale
of  energy. It is a phenomenological parameter which 
 does not appear in the Kirchhoff equations.

We emphasize that the Kirchhoff Hamiltonian is  only a part of the energy of the fluid. This part of energy is transported  by vortices. Another part of the  energy is related to the vortices at rest. It  diverges at vortex cores. This part is omitted in \eq{36}.

The Kirchhoff vortex system is  readily   canonically quantized. We  replace the Poisson brackets by the commutators\begin{align}\label{C}
\ii\hbar \{\bar z_i,\,z_j\}_{P.B.}\rightarrow[\bar z_i,\,z_j]=2\ell^2\delta_{ij}.
\end{align}
The parameter  \(2\ell^{2}=\hbar/(\Omega m_*)\) has  dimension of the area. It is a phenomenological parameter arising in the quantization. We measure  it in units of    area per particle \(2\ell^2=\nu/(\pi\bar\rho) \). Dimensionless
number
\(\nu\) \eq{321} is a semiclassical parameter.
We will see that \(\nu\) is identified with the filling  fraction and \(\ell\) with the magnetic
length. 
 
The next step is the choice of states. We assume that states are  holomorphic
polynomials of \(z_i\). Then operators
\(\bar z_i\)  are  canonical momenta \be\la{r}\bar  z_i  =2\ell^2\p_{z_i}.\ee
Finally, we have to   specify the inner product. We impose  the \emph{chiral
condition}: operators \(\bar z_i\)  and \(z_i\) are  Hermitian
conjugated
\begin{align}\la{hermitte}
\centering{\text{\small chiral condition}:\quad \bar z_i=z_i^\dag.}
\end{align} This condition combined  with the representation \eq{r} identifies  the   space of states with  the Bargmann space \cite{Bargmann} (see also \cite{GMP}). That is \ the Hilbert space
 of analytic polynomials \(\psi(z_1,\dots,z_N)   \) with the inner product
\begin{align}\la{B}
\langle\psi'|\psi\rangle=\int  d\mu \,\bar\psi'\psi,\quad d\mu=\prod_ie^{-\frac{|z_i|^2}{2\ell^2}}{d^2z_i.}\,
 \end{align}
 %In this representation  operators act on holomorphic functions. 
%Consequently \(\bar z_i^\dag=2\ell^2\p_{z_i}^\dag=-2\ell^2{\bar\p}_{ z_i}^\mathrm{T}+\bar
%z_i\).
Eqs. (\ref{u},\ref{r}) help to write down  quantum velocity  operators of vortices\begin{align}\la{v}
 &\pp_i=-\ii\hbar\left({ \p}_{z_i}-\sum_{j\neq
i}\frac{\beta}{z_i-z_j}\right),\\
&\pp_{i}=m_*\vv_{i},\quad {\dot {\overline z}_i=\vv_i, \quad \beta=\nu^{-1}}.
\end{align}
The operators \(\pp_i\) are the multi-particle version of operators of magnetic translations or guiding centers.

Eqs.(\ref{u}-\ref{v}) are quantum chiral Kirchhoff equations.
They are not difficult to be  generalized  to a sphere,  or a torus.
\section{ Quantum Chiral Kirchhoff Equations and FQHE}\la{V}The quantum  chiral Kirchhoff equations  are readily to identify with FQHE. 

The ground state of the  vortex  liquid is the state where the vortices are at rest. We repeat  that this state is  highly excited state   of the fluid. Rather it is a state of the fluid  at a very high angular moment. When vortices are at the ground state, the fluid moves with very high energy.

The ground state is  the analytic function whose phase is nulled by all momenta
operators. The common solution of  the  set of  1st order PDEs  $$\pp_i\psi_0=0$$ in the class of holomorphic polynomials is  the Laughlin w.f. in the Bargmann representation   \begin{align}\la{L} \psi_0(z_1,\dots,z_N)=\prod_{i>j}(z_i-z_j)^\beta,\quad \beta=1/\nu.\end{align}   
The wave function  is single valued if \(\beta\) is integer, antisymmetric if
\(\beta\)  is an  odd integer, symmetric if \(\beta\) is even integer. We see that vortices could be fermions or bosons depending of the choice of $\beta$.   In particular, if $\beta=2$ the same hydrodynamic equations describe the rotating trapped Bose.
 We notice how naturally the Laughlin wave function emerges in quantum hydrodynamics, both for fermionic and bosonic  FQH states.

 The correspondence will be completed if we assign  the electronic charge to vortices and to identify the angular
velocity with the effective cyclotron frequency\begin{displaymath}\Omega = \frac{eB}{m_*c}=\frac{m_b}{m_*}\omega_c\end{displaymath}The hydrodynamic interpretation of the FQHE  is subtly  different from  Laughlin's original interpretation. There, the coordinates entering the Laughlin w.f. are interpreted as bare band  electrons. The fluid itself   is absent in the Laughlin picture. The hydrodynamic  interpretation suggests that electrons (and their charge)
are localized on topological excitations (vortices) of a neutral incompressible  fluid.
The neutral fluid is real. It serves as the agent of the interaction between electrons. 

In the hydrodynamic  interpretation, the    quasi-hole
\cite{L} is a hole in the uniform  background of vortices.
It corresponds to a state  \eq{Q1} characterized by the   polynomial  with simple zeros located at  a given point \(z\) \begin{align}
Q(z_1,\dots,z_N)=\prod_i^N(z-z_i).
\end{align} The momentum  of this state is \(\pp_i|Q\rangle=\ii\nu \frac{\Gamma}{z_i-z}|Q\rangle\). It shows that the Magnus force between vortices and a quasi-hole is the opposite to    the  fraction \(\nu\) of the Magnus forces between vortices.  Thus  in the hydrodynamic interpretation   the quasi-hole appears  as a vortex  with a fractional negative circulation \(-\nu\) .

Identifying  vortices and electric charges, we must assume that  the   external
fields (potential well,  gradients of temperature etc.,) are coupled to  the vortices, not to the fluid. 

Let us examine how vortices move in the external potential well \(U(r) \). 
The potential  adds the term    \(\sum_i U(r_i)\) to the Hamiltonian, where \(r_i\) are coordinates of
vortices, and  adds the force   \(-i[U,\bar z_i]=
i2\ell^2\p_{z_i} U \) to the  Kirchhoff equations \begin{align}\la{44}\pp_ i=-\ii\hbar\left( \p_{z_i}- \sum_{i\neq
j}\frac{\beta}{z_i-z_j}\right)+\ii m_*\ell^2e\mathrm{E}(z_i),\end{align}
where \(eE=-\nabla U\) is the
electric field. 

Fractionally quantized Hall conductance  follows from Kirchhoff equations easily.
Let us assume for now that the electric field is uniform. Then  the center of mass of the fluid  stays at the origin \(\sum_i \p_{ z_i}=0\). Summing \eq{44} over all vortices we obtain  the  Hall  current per particle  \( N^{-1}\sum_ie\vv_i=\ii e^2\ell^2\mathrm{E}
   \) and the current per volume  \(\ii
e^2\ell^2\bar\rho\mathrm{E}\). We read that   the  Hall conductance
equals to the fraction of \(e^2/h\)\begin{align}\la{451}\sigma_{xy}=\nu\frac{ e^2}{h}.\end{align} 
%We discuss the Hall conductance in a non-uniform electric field in Sec. \ref{D}.
\bigskip
\begin{center}
\line(1,0){100}
\end{center}
\bigskip

 \noindent Our next step is  to  develop   the hydrodynamics
 of a system of quantum vortices described by the Kirchhoff equations. To the best of our knowledge this has not been done
even for the classical fluids. 
We start  by the summary of main results. The derivation and details follow.
\section{Summary of the main results and Discussion}\la{main}

Quantum hydrodynamics of the chiral vortex flow consists of
three sets of data: the operator content and their algebra, the chiral constituency relation between operators and the dynamic equation.
We summarize them below, but
first we comment on notations.

\subsection{Notations}

We will  use holomorphic coordinates  
 \(z=x+\ii y,\;\p=\frac{1}{2}(\nabla_x-\ii \nabla_y) \). We use
the roman script for complex vectors. For example, the velocity of the fluid 
 $$u=(u_x,u_y),\quad \mathrm{u}=u_x-iu_y.$$
 We denote the velocity of the vortex fluid
$$\mathrm{\vv}=v_x-iv_y,$$the 
momentum flux for the vortex fluid
$$ \mathrm{P}=P_x-iP_y.$$
%and holomorphic components of symmetric flux tensor \( 
%\Pi_{ab}:\) $$\Pi=\Pi_{xx}-\Pi_{yy}-2\ii\Pi_{xy},\;\Pi_{z\bar
%z}=\Pi_{xx}+\Pi_{yy}. $$
 
 We emphasize the difference between  Hermitian conjugation \(\vv^\dag\) and  complex  conjugation \(\bar \vv\), but still may use the classical   notation for the divergency and the curl of the velocity. In particular, the divergency and the curl we abbreviate as  \(\nabla\cdot v=0 \) actually means  \(\nabla\cdot v=\bar\p\vv+\p\vv^\dag,\;\nabla\times v=\ii(\bar\p\vv-\p\vv^\dag\)). Similarly,  the term \(v\cdot\nabla\rho\) in \eq{E1} is understood as
\(\vv^\dag\cdot\p\rho+\bar\p\rho\cdot  \vv\).  

The divergence-free velocity of the incompressible  liquid is expressed through the stream function operator  \begin{align}
 \vv=-2\ii\p\Psi.
\end{align}
We define the momentum flux of the vortex  flow as \begin{align}\la{P1}
\mathrm{P}=m_*\rho\vv. 
\end{align}
 The vortex flux operators  null the ground state
\begin{align}\la{254}\mathrm{P|0\rangle=\langle0|\mathrm{P}^\dag=0}. \end{align}
Throughout the paper we set \(m_*=1\) measuring the momentum per particle  in units of units of velocity, or equivalently, treating the particle density as a mass density. We emphasize that  \(m_*\) is not related to the band
electronic mass.

\subsection{Commutation relation}\la{BB}
Commutation relations of the vortex flux  operators differ from the
canonical commutation relations of quantum hydrodynamics  of Landau \cite{Landau} by
the anomalous terms

\begin{align}
%&[\mathrm{u}(r),\,\rho({ r'})]=-i \hbar \p\delta(r-r'),\\
%&
%&\text{\small Chiral condition}:\quad \Gamma^{-1}(\nabla\times u)=-{2\pi\mathrm{J}
%}(
%\rho -\bar\rho)+ j \Delta\log\rho,\\ 
%&\text{\small Heisenberg algebra}:\quad \Gamma^{-2}[\uu^\dag( r), \mathrm{u}(
%r')]=2j\pi\delta(r-r').\\
%&\frac{1}{2\hbar}[\mathrm{J}(r),\mathrm{J}^\dag(r')] = [-  J \times  %\nabla + 
%2\pi q\rho^2]\delta(r  -  r'),\la{JJ}\\
&\hbar^{-1}[\mathrm{P}(r),\mathrm{P}^\dag(r')] =-\frac 12( P \times   \nabla)\, \delta(r  -  r')+\underbrace{\frac{ \hbar}{2\nu}\left( 
{2\pi}\rho^2\delta(r  -  r')+\frac{1}{4} \nabla\left[ \rho\cdot\nabla\delta(r-r')\right]\right)}_{\text{
anomalous term}}.\la{PP1}
\end{align}The commutation relation between the flux and the density  is canonical
\begin{align}\la{91}
[\mathrm{P}(r),\rho(r')]=-\ii\hbar\rho\p\delta(r-r').
\end{align}The vortex flux operator can be conveniently represented in terms of canonical fields \(\uu\) and \(\uu^\dag\)\begin{align}
[\uu(r),\uu ^\dag(r')]=-2\pi\frac{\hbar^2}{\nu}\delta(r-r'),\quad [\mathrm{u}(r),\rho(r')]=-\ii\hbar\p\delta(r-r')
\end{align} We introduce the axillary operators 
\begin{align}\la{12}
\mathrm{J}=\rho\uu\quad 
\end{align}
which we call vorticity
flux. The hydrodynamic interpretation of this operator will be clarified in the text. It has a canonical commutation relation with itself and with the density, but does not annihilate the vacuum. The  vortex flux \({\mathrm P}\ \) does (as in \eq{254}).  

We show that vortex flux and  vorticity  flux 
are differed by  the anomalous term   
 \begin{align}\la{p1}
 \mathrm{P}=\mathrm{J}+\ii\frac{\hbar}{2\nu}
 \p\rho.
\end{align}
 The  anomalous term adds to the
diamagnetic  energy of the flow in the background e.m. field\\ \begin{displaymath}\frac{e}{c}\int (A\cdot P   )d^2
r=\frac{e}{c}\int
(A\cdot J)  d^2
r+\frac{\hbar}{4\nu}\frac{e}{c}\int
B\,\rho d^2r,\end{displaymath}effectively  reducing the orbital moment of particles.
 Similarly, the anomalous term contributes to the angular momentum of the flow  \begin{displaymath}N^{-1}\int(r\times P) d^2r=N^{-1}\int(r\times J) d^2r+\frac{\hbar}{4\nu}\end{displaymath}The  meaning of the anomalous term   is  seen directly from the monodromy of  FQH
states \eq{Q1}.
Monodromy with respect to a closed path  is the phase
acquired by the wave function  when a particle is moved   along  that path. That
is   a  circulation  of the  particle. It equals to the number
 of zeros  of the wave function with respect
to each coordinate. This number is \((n-1)/\nu\), where \(n\) is the number of particles  enclosed by the path. It is less by \(\nu^{-1}\)
from the number of magnetic flux quanta
piercing the system, simply because the vortex does  not interfere with itself.
The  anomalous term  accounts for that difference.  
 The anomalous term  can be seen as a local version of the global relation
between the monodromy of states and the number of particles. The difference,
often called   the shift \(2\bar s\), has been emphasized  in Ref. \cite{shift2}.
For the  Laughlin states \(2\bar s=\nu^{-1}\).

\subsection{Anomalous term in the chiral constituency relation}\la{A}

Unlike in a regular fluid mechanics where the density \(\rho\) and velocity \(v\) are independent fields in the chiral flow they  are related by
the {\it chiral constituency relation}. This means that the set of states on the lowest Landau level is restricted in such manner that the velocity operator
acts
 as a certain functional of the density operator. 

In a very rough approximation the chiral relation
states that the vorticity  per particle is the inverse filling factor in units of the Planck constant  as  suggested  in \cite{Stone}.
This view refers to a popular picture of the FQH states as electronic states with an additional amount of flux  attached to each particle.  The actual relation between the vorticity and the density  is more complicated.  It involves  the {\it anomalous term}
\begin{align}
   \nabla\times v= \frac{h}{\nu}[\rho-\bar\rho+\underbrace{\frac{ 1}{4\pi}(\frac{1}{2}-
\nu)\Delta\log\rho}_{\text{anomalous term}}],\la{v3}\end{align}
 where \(\bar\rho=\nu(2\pi\ell^2)^{-1}=\nu\frac{e}{h c}B\) is the mean density of electrons, and \(h=2\pi\hbar\).

An accurate reading of this relation is: the action of the operators in the r.h.s and the l.h.s  of \eq{v3} on the Bargmann "bra" state \(\langle Q|   \) are equal. They are not equal if the "bra"-state is not in the Bargmann state.

In particular, a quasi-hole, a source
for vorticity localized at \(r_0\), corresponds to the polynomial \(Q=\prod_i(z_0-z_i)\). It deforms the density outside of the core \(r=r_0\) according to the equation   \cite{comment1}
\begin{align*}-\nu\delta(r-r_0)= \rho-\bar\rho+\frac{ 1}{4\pi}(\frac{1}{2}-
\nu)\Delta&\log\rho. \end{align*}
An equivalent  form of  the chiral relation connects the stream function and the density
\begin{align}\la{7}
v_a=-\epsilon_{ab}\nabla_b\Psi,\quad \Psi= \frac{\hbar}{2\nu}[\varphi-(\frac{1}{2}-
 \nu)\log\rho],
\end{align}
where  the "regular part" of the stream function
\(\varphi\)  
%\(\varphi=-2\sum_i\log|z-z_i| +\pi\bar\rho{|z|^2}-j\log N+j\p_N\), 
is a solution of  the Poisson equation \begin{eqnarray} \la{9}\Delta
\varphi = - 4\pi( \rho - \bar\rho).\end{eqnarray}
% Another suggestive form of the chiral relation reads
% \begin{align}
% &\mathrm{P}=\frac{\ii \hbar}{\pi\nu}[-\pi\bar\rho\p\varphi+\bar\p\mathcal{T}],\quad
% \mathcal{T}=\frac{1}{2}(\p\varphi)^2-\underbrace{(\frac12-\nu)\p^2\varphi}_{\text{anomalous
% term}}.\la{v4}
% \end{align}
We comment that the chiral relation readily to extend to the case of the inhomogeneous magnetic field. In this case the mean  density \(\bar\rho=\nu\frac{e}{h c}B\) in \eq{v3} and \eq{9} is the function of coordinates.
There is no other changes.
In particular, in the ground state, where the velocity vanishes, the  density in a non-uniform magnetic field obeys the "Liouville equation with the background". That is  \eq{v3} with zero in the l.h.s. In the leading order in gradients, the ground state density acquires a universal correction
\begin{align}\la{341}
\rho = \nu\frac{e}{h
c}B-\frac{ 1}{4\pi}(\frac{1}{2}-
\nu)\Delta&\log B+ \dots
\end{align}
 The  integrated form of 
\eq{v3} is   the  sum rule connecting the angular momentum \(\mathrm{L}=(\hbar^{}N)^{-1}\int(  r \times  
 P)d^2r\) (per particle in units of \(\hbar\))  to the gyration per particle  
\(N^{-1}\mathrm{G}=\int  r^2(\rho-\bar\rho)d^2r\) of the flow\begin{align}
\ell^2\mathrm{[L-(\frac 1\nu-2)]}=
\mathrm{G}.
\end{align}
The ground state version of this formula is the familiar sum rule of the Laughlin w.f. \begin{displaymath}\nu\langle0|\sum_i|z_i|^2|0\rangle=\ell^2N(N-1+2\nu).\end{displaymath}.

\subsection{Anomalous term in the Euler Equation: Lorentz shear stress}\la{CC}The constituency relation \eq{P1}, the chiral  condition \eq{v3},  the continuity equation \eq{E1}, and
the operator algebra (\ref{PP1},\ref{91}) is
 the full set
of hydrodynamics equations for the chiral incompressible quantum fluid. 

The  chiral condition  helps to write the continuity equation \eq{E1} as
a  non-linear equation of  the density alone\begin{align}\la{17}
\p_t\rho-\frac{\hbar}{2\nu}\nabla\varphi\times \nabla\rho=0,\quad \Delta
\varphi = - 4\pi( \rho - \bar\rho) 
\end{align}
The equation is identical to the Euler equation for the vorticity in the  incompressible fluid. Naturally the anomalous term disappears from this equation. It   appears in the boundary conditions, in  the response to external fields and, also, determines forces acting in the fluid. 

Forces are rendered by the    momentum flux tensor    \(\Pi_{ab}\), entering the  Euler equation, written  in the form of the conservation  law  \begin{align}\la{T}
%\p_t {\mathrm{\mathrm{P}}}=\p \mathrm{ T}_{z\bar z}+\bar\p\mathrm{ T}_{zz},
\p_t P_a+\nabla_b \Pi_{ab}=\rho F_a.
\end{align}
 Here      \(  F=eE-\frac
ec B  \times v\) is the Lorentz force. 

The \emph{anomalous viscous stress} emerges in the momentum stress tensor. A general  fluid momentum flux tensor of incompressible fluid consists of the kinetic part, the stress,  and the traceless viscous
stress \(\sigma'_{ab}\). In the incompressible fluid the stress is expressed through the velocity. We write \(\Pi_{ab}=\pi_{ab}-\sigma'_{ab}\), where \(\pi_{ab}\) accounts for the kinetic part and the stress. At a fixed density \(\pi_{ab}\) is  symmetric with respect to a change of the direction of the velocity \(v\to -v\). The viscous term  is linear in gradients of  the velocity. It changes the sign under this transformation. With the exception of the diamagnetic term, the viscous term has  a lesser degree of  velocity among terms of the flux tensor. This is the 
only term enters the linear response theory. 

Our fluid is dissipation-free. Therefore the anomalous viscous stress produces no work. This is possible if the viscous stress represents forces acting normal to a shear. Such stress  could only be  a traceless pseudo-tensor. It changes sign under the spatial reflection. In the vortex flow  the anomalous viscous stress reads
\begin{align}\la{vis}
\quad \sigma_{ab}'=-\frac{\hbar}{2\nu}\bar \rho(\nabla_a\nabla_b-\frac
12\delta_{ab}\Delta)\Psi.
 \end{align}
% In holomorphic coordinates
% this equation reads \(\p_t\mathrm{P}+\p\Pi_{\bar z z}+\bar\p\Pi=\rho \mathrm{F}\), where 
% \(
% \Pi=\Pi_{xx}-\Pi_{yy}-2\ii\Pi_{xy}\), holomorphic component of the tensor, and \(\Pi_{z\bar
% z}=\Pi_{xx}+\Pi_{yy}
% \)
% is  its trace.
 There is a noticeable difference with the dissipative shear viscous stress. That stress is given by the same formula  where    the  stream function is replaced by the hydrodynamic potential. 

Components of the anomalous viscous     stress tensor are
 \begin{align}
\sigma'_{xx} = -\sigma'_{yy}  =-\frac{
\hbar }{4\nu}\bar \rho(\nabla_xv_y + \nabla_yv_x),\;\sigma'_{xy} = \sigma'_{yx} = \frac{
\hbar }{4\nu}\bar\rho(\nabla_xv_x - \nabla_yv_y)
\end{align}
 The divergency  of the Lorentz shear stress is the Lorentz shear force \begin{displaymath}\nabla_b\sigma_{ab}'=\frac{
\hbar }{4\nu}\bar\rho\nabla_a(\nabla\times v)\end{displaymath} exerted by the  flow on the volume element of the liquid.   It is proportional to the gradient of the vorticity. 

A noticeable feature of the anomalous stress is that the kinetic coefficient \(\frac{
1 }{4\nu}\) (in units of \(\hbar\)) is universal and has a geometrical origin. The anomalous conservative viscosity is  referred as  odd-viscosity, or Hall viscosity. It  has been introduced in Ref. \cite{Avron} for the Integer Hall Effect as a linear response to a shear. Its notion has been extended to the FQHE in \cite{ReadHV,Tokatly} (see \cite{HV} for an incomplete list of references).
In this paper we show how  the anomalous viscosity appears in the  hydrodynamics of the chiral flow.

The anomalous term \eq{vis} represents the force acting  normal to the shear (in contrast, the shear viscous force acts in the direction parallel and opposite to the shear). The anomalous viscous stress is also referred as the Lorentz stress, and the force is referred   as the Lorentz shear force \cite{Tokatly}. 

Emergence of the Lorentz shear stress can  be interpreted in terms of semiclassical
motion of electrons.
A motion of  electrons consists of a fast motion along  small orbits  and
a slow motion of  orbits. A shear 
flow strains orbits elongating them normal to the shear, boundaries and vortices.
 Elongation
yields    the Lorentz shear stress. 

\subsection{Topological sector}\la{EE1}
The topological sector consists of flows driven by slow long-wave external fields, such as curvature of the space,  a non-uniform
electric and magnetic fields, etc., which does  not produce excitations over the gap. The fluid may be driven, but remains in the ground state. Hall current is the most familiar example. The flow in the topological sector is  steady.  

The topological sector can be singled out in the limit \(\Delta_{\nu}\to\infty\). In this limit  the momentum flux tensor reduces to the  anomalous viscous stress modified by the quantum corrections. Then  the   dynamics
reduces to the balance between the  Lorentz shear force and the Lorentz force.

  In the linear approximation the stationary Euler equation reads
\begin{align}\la{22}
 \left( \frac{1}{4\nu}-\frac 12\right)   \nabla (\nabla\times v)=e  E-\frac
ec B\times v.  
\end{align}
Solution of the is equation in  the leading gradient approximation yields  the universal correction to the Hall conductance   \cite{Son} 
  \begin{align}\la{23}
\sigma_{xy}(k)=\frac{\nu e^2}{h}\left(1+ (\frac{1}{4\nu}-\frac 12)(k\ell)^2\right) .\end{align}
The Hall current increases with the wave-vector.

 The factor $1/2$ in these equations represents the diamagnetic energy. This term does not
appear explicitly  in the quantum equation  \eq{T}. Rather it is hidden  in
the normal ordering of the kinetic part of the vortex flux tensor.  If in
addition, particles possess an  orbital moment \(M\) which is intrinsically related to the band,  the  term \(\frac{m_*}{m_b} M\) is added to the factor \(-1/2\) in both  equations. Apart from  this effect the correction to the Hall conductance is universal.

The universal correction to the Hall conductance can be also seen directly from the Eq. \eq{341} describing the density in a non-uniform magnetic field. The Hall conductance connects the density and magnetic field in the linear approximations (the Streda formula):  \(e\rho_k=\frac 1 c\sigma_{xy}(k)B_k\).

\subsection{Trace and mixed anomaly} 

The  meaning of the Lorentz shear stress is best illustrated when 
the fluid is placed into a curved space. In this case the energy  receives
an addition \( H'=\frac{\hbar}{4\nu}\bar\rho\int   R\Psi\sqrt g d^2\xi\),
where \(R\) is the spatial scalar curvature. This addition yields the 
 \emph{trace anomaly}: the flux tensor acquires a trace   proportional to the curvature\begin{align}
-\sigma'_{aa}=\bar\rho\frac{\hbar^2}{16\pi\nu} R. \quad 
\end{align}
 It is traceless if the space is flat. This formula experiences quantum corrections, effectively shifting \(\nu^{-1}\) by a number. 
The trace anomaly yields to a uniform force acting toward the region with the access curvature. This force squeeze particles toward the curved regions
(\emph{mixed anomaly})\begin{align}
\delta\rho=\frac{1}{8\pi}R\sqrt g.
\end{align}Accumulation of charges at curved parts of space has been
suggested in \cite{shift2} and further discussed in \cite{Son}.

 \subsection{Boundary double layer and dispersion of edge modes}A striking manifestation of the anomalous terms is seen on the boundary. The Lorentz shear force squeezes flow lines with different velocities. As a result the charge there is an accumulation of density on the edge.
The density at the edge \(r=R\)  forms the \emph{double layer}  \begin{align}
\rho(r)=\bar\rho+\frac{1-\nu}{4\pi}\,\nabla_n\delta(r-R).
\end{align}
Here the derivative is taken in the direction normal to the boundary.

A consequence of the double layer is the correction to the spectrum of edge modes
\begin{align}\omega(k)=c_0k+\frac 12\Delta_\nu\left(\frac{1}{\nu}-1\right)\,{\rm{sign}}(k) (k\ell)^2,
\quad c_0=c\frac{E}{B}.
\end{align} 
These results where obtained in \cite{W2}.
% These data could be combined by the extended  chiral relation \eq{7} by including the metric (spin connection \(\epsilon_{ab}\nabla_\nu\omega_a=\frac 12R\sqrt g\)) and electric field
% \begin{align}
% v_a=u_a+\frac{\hbar}{4\nu}\epsilon_{ab}(\nabla_\nu\log\rho-\frac{1}{\pi}\omega_\nu)+\frac{\hbar c}{B}\epsilon_{ab}E_\nu
% \end{align}
% 
\bigskip
\begin{center}
\line(1,0){100}
\end{center}
\bigskip In the rest of the paper we obtain these (and some other) properties starting from quantized chiral fluid.
   It happens that  many calculations are merely identical in the classical and the quantum cases. To simplify the matter, we first   derive
 the hydrodynamics of the vortex fluid in the classical case, and then consider the quantum case. 
\section{Relation between velocity of the vortex flow and velocity of   the fluid.}\la{R}
\noindent Eulerian hydrodynamics
of the vortex flow describes the flow in terms of  the  density and velocity field 
\(v(r)\) of vortices. We construct the velocity field starting from velocities of individual vortices.  The calculations
are merely
identical  in the classical and the quantum cases. We proceed with the classical calculations.

\bigskip

\noindent 
 For the purpose of the hydrodynamics of the vortex flow we denote the density of vortices   as \begin{align}\la{45}
 \rho(r)=\sum_{i}\delta(r-r_i)=\bar\rho+\frac{1}{2\pi\Gamma}(\nabla\times
u).
 \end{align} 
Having the density one finds the velocity, or the stream function of the fluid. The stream function of the fluid is  the potential \(\varphi\) \eq{9}\begin{align}\la{49}
\uu=-2\ii\Gamma\p\varphi,\quad \Delta\varphi=-4\pi(\rho-\bar\rho).
\end{align} The object of interest is the  vortex flux  \begin{align}\la{P}
\mathrm{P}(r)=\sum_{i}\delta(r-r_i)\vv_
i.
\end{align}
  Having the flux, we define the velocity field of the vortex fluid as
\(
\mathrm{P}= \rho\vv.
\) 
We want to compute the velocity of the vortex flow \(\vv(r)\), and to compare it with the velocity of the original fluid \(\uu(r)\). Obviously, they are  different. The former describes the slow motion of vortices, the latter the fast motion of the fluid around vortices and the drift together with vortices. Nevertheless there is a simple relation between the two. 

We compute the  vortex  flux \(P\) and compare it with the vorticity flux
\(
\mathrm{J}= \rho\uu,
\)
where the velocity of the fluid \(\uu\) is  given by \eq{k}.
\bigskip

Using
\eq{u} and the  \(\bar\p\)-formula
\(\pi\delta=\bar\p(\frac 1z)    \) we write
\begin{align}\la{35}
\mathrm{P}(r)=\sum_i\delta(r-r_i)[-\ii \Omega\bar z_i+\ii\sum^N_{i,i\neq j
}\frac{\Gamma}{z_i-z_j}]=-\ii \Omega\bar z\rho(r)+\ii\frac{\Gamma}{\pi}\bar\p\sum^N_{i\neq
j
}\frac{1}{z-z_i}\frac{1}{z_i-z_j}.
\end{align}
Then use the identity \begin{align}
2\sum_{i\neq
j}\frac{1}{z-z_i}\frac{1}{z_i-z_j}=(\sum_i\frac{1}{z-z_i})^2-\sum_i(\frac{1}{z-z_i})^2 =(\sum_i\frac{1}{z-z_i})^2+\p\sum_i\frac{1}{z-z_i}
\end{align} 
and apply \(\bar\p\)
\begin{align}
\rho\vv=-\ii \Omega\bar z\rho+{\ii \sum_i\delta(r-r_i)}\sum_j\frac{\Gamma}{z-z_j}+\ii\frac \Gamma 2\p\sum\delta(r-r_i).
\end{align}
We obtain the relations 
 \begin{align}\la{shift3}
  \mathrm{P}=  \mathrm{\rho\uu}+\frac {\Gamma}{2}  \ii\p\rho,\quad    \vv=  \uu+\frac
{\Gamma}{2}\rho^{-1} \ii\p\rho.
\end{align}
  The difference between the velocity of the vortex fluid and the velocity of the fluid has a simple meaning. Velocity of  the fluid \(u\) diverges at
the core of an isolated vortex
 (as it is seen   in (\ref{k})). However, velocities of  vortices are finite.
The  anomalous term  removes that singularity.

The anomalous term  changes only the transverse part of the velocity, so that  the flow of vortices is  incompressible
 like the fluid itself \(\nabla v=\nabla u=0\).
Also, the anomalous term does not change the divergency of the flux \(\nabla P=v \nabla\rho=\nabla(\rho u)=u \nabla\rho\).

\section{Classical Hydrodynamics of the Chiral  flow}\la{Q}

%\subsection{Anomalous vortex flux tensor: Lorentz shear stress}\la{T2}

Global symmetries of space and time, such as translation and rotation yield familiar conservation laws of the flux, energy and angular momentum.  In addition, the 2D  incompressible flows with a constant density possess conservation laws which are not directly related to  global symmetries. One conservation law  is familiar. This is the conservation of vorticity.  With the help of  (\ref{shift3}) the   Euler equation in
the form of  \eq{E} appears as the continuity  equation for the mass density
of the vortex fluid
\begin{align}\la{54}
D_t\rho\equiv\left(\frac{\partial}{\p t} +
  v\cdot \nabla\right)\rho=0.
\end{align}  In addition to conservation of vorticity, the vorticity flux \(J\) and the vortex flux \(P\) are also conserved\begin{align}J=\rho u,\quad P=\rho v.\end{align}
Conservations of  the vorticity and    the vortex flux  are obvious in the Kirchhoff picture. They are conservation of mass and mass flux of the vortex system. The vorticity flux is conserved due to the relation \eq{shift3}.  In continuum fluid mechanics
conservation of the  vorticity and vortex fluxes are less obvious.  They follow from the observation that the vorticity flux is the flux of the fluid plus  a divergency of the symmetric and traceless tensor
\begin{align}\la{541}
J_a=\bar\rho u_a+\frac{1}{2\pi\Gamma}\epsilon_{ab}\p_{ c} t_ {b c},\quad  t_ {b c}=u _b u_ c-\frac 12\delta_ {b c} u^2.
\end{align}
The  conservation law for  the vorticity flux introduces the vorticity flux tensor\begin{align}\la{57}
\p_t J_a+\nabla _b \pi_{ab}=-\frac
ec \rho(B  \times u)_a
\end{align}
  The r.h.s. of this equation is the Lorentz force.

The vorticity flux tensor can be locally and explicitly expressed through the velocity and the pressure. Expression is cumbersome and we do not need it for the purpose of this paper. In the leading approximation in the density gradients
the second term in \eq{541} could be dropped. Then  
%  consists of the kinetic part and the stress. In the case of the vorticity %flux both are traceless. Straightforward   calculations  combine  the Euler %equation 
% \(\p _tu_a+\nabla _b( u_a u _b+p\delta_{ab})=0\) and the continuity
% equation for the vorticity \eq{54}. They yield that the vorticity flux %\(J\) is conserved if one represents \(\rho\nabla p\) as a divergency of %a symmetric tensor \begin{displaymath}\nabla _b\sigma_{ab}=\rho\nabla_ap,\end{displaymath}which depends on  velocity and pressure locally.  Tedious calculations yield\begin{align}\la{591}
% \sigma_{ab}=\rho\, u_a u _b-(\epsilon_{a c}u _b\nabla_ c+\epsilon_ {b c}u_
% c\nabla_a)\left(\frac{|u|^2}{2}+
%  p\right).
% \end{align}We remind  that in an  incompressible flow velocity  determines the pressure, which can be excluded from \eq{591}\begin{align}
% \frac{|u|^2}{2}+
%  p=\frac{1}{4\pi}\int(J\times\nabla_z)\log|z-\xi|d^2\xi.
% \end{align}
%  As a result we obtain the vorticity flux tensor \begin{align}\la{572}
% \pi_{ab}=\rho\, t_{ab  }+\sigma_{ab}-\frac 12\delta_{ab}\sigma_{ c c.}
% \end{align}   It is traceless. For references we write
% the vorticity flux tensor  in the complex form and solely through the velocity\begin{align}
% \pi=2(\pi_{xx}-\ii\pi_{xy})=\uu\left(\rho\uu+\int\frac{\p\mathrm{\bar
% J}}{z-\xi}\frac{d^2\xi}{\pi}\right).
% \end{align}
 the vorticity flux tensor is identical to the flux tensor of the incompressible fluid with the constant density  \(\pi_{ab}\approx\bar\rho u_au_b+p\delta_{ab}\).

 The next step is to determine the vortex flux tensor \(\Pi_{ab}\). It enters   into the conservation law for the vortex flux \begin{align}\la{611}
\p_t P_a+\nabla _b\Pi_{ab}=-\frac
ec \rho(B  \times v)_a.
\end{align}
We see it  as a transformation of the vorticity flux tensor induced by the transformation of the velocity  \eq{shift3}\begin{align}
u\to v,\quad J\to P,\quad\pi\to \Pi.
\end{align}
Under the shift \eq{shift3} we have \(\dot P_a=\dot J_a+\frac{\Gamma}{4}\epsilon_{ab}\nabla _b\dot\rho\). With the help of the continuity equation \eq{54}, we obtain the transformation \begin{displaymath} \pi_{ab}\to \Pi_{ab}=\pi_{ab}+\frac{\Gamma}{4}[\epsilon_{ac}\nabla_ c(\rho v_ b)+\epsilon_{bc}\nabla_
a(\rho v_ c)].\end{displaymath} 
In the leading approximation in gradients we replace the density in the last equation by its mean \(\bar\rho\).
We observe that the stress tensor acquires the  anomalous  viscous term \begin{align}
\Pi_{ab}\approx\pi_{ab}-\sigma'_{ab}, \quad \sigma'_{ab}=-\frac{\Gamma}{4}\bar\rho(\epsilon_{a c}\nabla_ c v
_b+\epsilon_ {b c}\nabla_a v_ c).
\end{align}
This is the Lorentz shear stress
\cite{HV}.

We see that  the Lorentz shear stress naturally appears in  the vortex liquid.
Chiral flows   consist of a fast motion along  small orbits around vortex
cores  and a slow drift
of  centers of these  orbits. A shear 
flow strains orbits elongating them normal to the shear, boundaries and vortices.
 Elongation
yields     to the Lorentz shear  stress. 
% 
% Important feature of the anomalous term is that it emerges as a  dominant  in the leading gradient expansion. Its gradient is the   Lorentz
% 
% We present\begin{align}
% t_{ab}=u _a u_ b-\frac 12\delta_
% {ab } u^2\to T_{ab}=v _b
% v_ c-\frac 12\delta_ {b c} v^2+\frac{\Gamma}{4}(\epsilon_{a c}\nabla_ c v _b+\epsilon_ {b c}\nabla_a v_ c).
% \end{align} The transformation of the  stress in \eq{572} reads \begin{displaymath} \sigma_{ab}-\frac 12\delta_{ab}\sigma_{ c c}\to\sigma_{ab}=\sigma_{ab}-\frac 12\delta_{ab}\sigma_{ c c}-\left(\frac{\Gamma}{4}\right)^2\rho(\nabla_a\nabla _b-\frac 12\delta_{ab}\Delta)\log\rho+\frac 14\rho(\nabla\times v)\delta_{ab}\end{displaymath}
% we note that the stress acquires a trace. 
% 
% Summing up we write the vortex flux as a sum of the kinetic part and the stress\begin{align}
% \Pi_{ab}=\rho\, T_{ab}+\sigma_{ab}.
% \end{align} 
% and emphasize the anomalous term in the kinetic part of the flux tensor represents the anomalous viscous term, or the Lorenz shear stress
% \begin{align}
% \rho T_{ab}=\rho(v _b
% v_ c-\frac 12\delta_ {b c} v^2)-\sigma'_{ab},\quad \sigma'_{ab}=-\frac{\Gamma}{4}\rho(\epsilon_{a c}\nabla_ c
% v _b+\epsilon_ {b c}\nabla_a v_ c).
% \end{align}
% To conclude  we write  the  kinetic part
% of the vortex flux stress  in the complex form \begin{align}
% \mathrm{T}=\vv^2-2\ii\p\vv=-4[(\p\Psi)^2+\Gamma\p^2\Psi].
% \end{align}
% The anomalous term is the second term in the r.h.s.

\section{Quantum hydrodynamics of the vortex matter}\la{S}We start by  quantizing
the incompressible chiral 2D fluid and then proceed with  the quantization
of the   vortex flow.
\subsection{Quantum hydrodynamics of incompressible liquid}
 The  canonical fields in hydrodynamics are density and velocity. In the chiral  fluid  with a constant  fluid density the  canonical hydrodynamics variables are velocity \(\uu \) and  vorticity \(\rho\), or rather,  holomorphic and anti-holomorphic   components of the velocity \(\uu\) and \(\uu^\dag\). 

%It should be noted a subtlety  of quantizing hydrodynamics in the Bargmann space. The density \eq{45} is real and therefore consists of holomorphic and anti-holomorphic variables. We "decompose" it into the  holomorphic and anti-holomorphic
%parts using the \(\bar\p\) formula \(\pi \delta(r)=\bar\p(\frac{1}{z})=\p(\frac{1}{\bar
%z}) \) as \(\rho(r)=\rho_++\rho_-=\frac {1}{2\pi}\bar\p\sum_i\frac{1}{z-z_i}+\frac
%{1}{2\pi}\p\sum_i\frac{1}{\bar z-\bar z_i}\). In the Bargmann space the action of the holomorphic  operator \(2\p_{z_i}\) acting  on the density is not just a differentiation over coordinates \(\p_{x_i}-\ii\p_{y_i}\) as it may appear from the notation. It acts only on the holomorphic part \(\rho_+\). Hence \(2\p_{z_i}\rho=-\p\delta(r-r_i)\) is half the regular derivative.
%We already encountered this subtlety  in Sec. \ref{C1} discussing the action of velocity in the "first quantized" formalism. 

Bearing this nuance in mind the quantization of the fluid velocity amounts to the replacement  of the term \(-\ii\Omega \bar z\) in \eq{k} by \(\p\pi_\rho\) ,  where  
\(\pi_\rho=-\ii\hbar\frac{\delta}{\delta\rho}\)  is the canonical momentum the density.
Also we  
replace the sum in  \eq{k} by the integral \(\sum_i\frac{\Gamma}{z-z_i}\to\Gamma\int
\frac{\rho(\xi)}{z-\xi}d^2\xi=-\ii\frac\hbar\nu\p(\varphi+\pi\bar \rho |z|^2)\). 
  We obtain  the velocity of the quantum chiral fluid  \begin{align}\la{53}
\uu=\p\left(\pi_\rho- \ii\frac\hbar\nu(\varphi+\pi\bar \rho |z|^2)\right).
\end{align}
This formula yields  the canonical commutation relation
between vorticity and velocity and between components of velocity\begin{align}\la{555}[\uu(r),\rho(r')]=-\ii\hbar\p\delta(r-r'),
\quad\nabla\times u=\ii(\bar\p\uu-\p\uu^\dag)= \frac{h}{\nu}(\rho-\bar\rho).\end{align}
The commutation relations between components of velocity  is the canonical
Heisenberg algebra as it is known to be in a quantizing magnetic field   \begin{align}\la{571}
[\uu(r),\uu^\dag(r')]=-2\frac{h^2}{\pi\nu}\delta(r-r'),\quad [\uu(r),\uu(r')]=0.
\end{align}

The algebra is completed by the equal point commutator\begin{align}
[\uu(r),\rho(r)]=\ii\hbar\p\rho(r)\la{701}.
\end{align}

The remaining element of the quantization is the  chiral condition. The holomorphic derivative \(\p_{z_i}\) acting to the left on  the
anti-holomorphic "bra"
states of the Bargmann space \eq{Q1} differentiates only the  factor
\(\exp({-\sum_i\frac{|z_i|^2}{2\ell^2}})\) of the measure \(
\langle Q|\left
(2\ell^2\p_{z_i}^\mathrm{T}+\bar z_i\right )=0.
\)
 Similarly the operator \(\p\pi_\rho\) acting on the left acts only on the factor \(\exp (-\frac{1}{2\ell^2}\int \rho
d^2r)\)
  \begin{align}\la{52}
 \langle
Q|\left (\p\pi_\rho+\ii\frac{\hbar}{\ell^2}\bar z\right )=0.
\end{align}
Therefore, when %\subsection{The chiral condition for the chiral fluid}
  the holomorphic velocity operator acts on the ant-holomorphic "bra" state  the first two terms in \eq{53} cancel. We return to the classical formula \eq{49}\begin{align}\la{702}
\langle Q|\uu+\ii\frac\hbar\nu\p\varphi|Q'\rangle=0.
\end{align}We emphasize that this relation does not hold if the operator is not sandwiched between anti-holomorphic and holomorphic states. 

The  chiral condition projects all operators onto the lowest Landau level. The projected velocity is manifestly divergence-free. Projection onto the lowest Landau level is summarized by the condition   \(\Delta\pi_\rho=-4\pi\bar\rho\).

The Heisenberg algebra of velocities \eq{571},  the continuity equation
 for the vorticity \(D_t\rho=0\) \eq{54}, and the chiral condition \eq{702} summarize the quantization of  hydrodynamics
of incompressible chiral flow. 
\bigskip\bigskip

Finally we are ready to  proceed with the quantization of the vortex fluid.

\subsection{Quantization of the vortex fluid}\la{T1}
The classical formula for   the flux  \eq{P} must be treated as an ordered  product of operators\begin{align}\la{704}
\mathrm{P}(r)=\sum_i\delta(r-r_i)\pp_i=\sum_i(\pp_i+\ii\hbar\p_{z_i})\delta(r-r_i),
\end{align} 
where the momenta \(\pp_i\) are given by \eq{44}. The relation  between the  velocity of the fluid \(\uu\) and the velocity of teh vortex fluid \(\vv\) \eq{shift3} holds on the quantum level \begin{align}\la{61}
 \mathrm{\mathrm{P}}=\rho \uu+\ii\frac{\hbar}{2\nu}
 \p\rho. 
 \end{align}
   The chiral  condition is obtained by placing \(\uu\)  to the left. Using \eq{701}, or equivalently   \eq{704},   pull  \(\uu\)  to the left 
and reduce it to its classical value \eq{49}.   This yields   the chiral conditions
of Sec. \ref{A}
\begin{align}\la{721}
\mathrm{P}=-\ii\frac\hbar \nu\rho\p\varphi+\ii\hbar(\frac{1}{2\nu}
 -1)\p\rho.
\end{align}The commutation relations of flux components presented in Sec.\ref{BB} by Eqs. (\ref{PP1},\ref{91}) follow.

Computation of the quantum vortex  flux tensor is not much different from the classical version of the Sec.\ref{R}. All the formulas remain the same provided that normal ordering of operators is respected. 
However, when the
velocity  is pulled to the left the coefficient in front of the Lorentz force
acquires the quantum correction \(\frac{1}{2\nu}\to\frac{1}{2\nu}-1\).

\section{Applications}\la{X}
\subsection{Structure function}
Anomalous  commutation relations help to compute the structure factor. This is  \(s_k=N^{-1}\langle0|\rho_k\rho_{-k}|0\rangle\), where   \(\rho_k
%=\int e^{-ikr}(\rho-\bar\rho)d^2r
 =\sum_i^Ne^{\ii k\cdot r_i}\) is the Fourier mode of a small density  modulation with the wave vector \(k\). 
%It is proper to think about the spectral function as a square of the norm %of a plane wave \(|k\rangle=\sum_i^Ne^{\ii k\cdot r_i}|0\rangle\). The spectral %function is \(\langle0|\rho_k\rho_{-k}|0\rangle=\langle k|k\rangle\). 

The chiral condition connects modes of density  and  flux. Let us evaluate it in  the linear approximation in density modes. Using \(k^2\varphi_k=4\pi\rho_k\) in \eq{721} we write the Fourier mode of the  flux 
%\begin{displaymath}\mathrm{P}_k=\int \mathrm{P}(r) e^{\ii k\cdot r} d^2 r=\frac{1}{2}\sum_i^N\left(e^{\ii k\cdot r_i}\vv_i+\vv_ie^{\ii k\cdot r_i})\)
 in terms of density modes  \begin{align}\la{74}
\mathrm{P}_{k }= \frac{\rm \hbar k}{(\ell k)^2}\left(1 - \frac{1}{2}(\frac{1}{2\nu}-1)(k\ell)^2\right )\rho_k,\quad {\rm k}=k_x-\ii k_y.
\end{align}
On the other hand the commutation relation
\eq{91} reads
\begin{align}
[\mathrm{P}_k,\, \rho_{-k} ] =\frac 12N\hbar {\rm k}.
\end{align}
 Since \(\mathrm{P_k}\) annihilates the ground state \(\langle0|[\mathrm{P}_k,\,\rho_{-k} ] |0\rangle=\langle0|\mathrm{P}_k\rho_{-k}|0\rangle\). We obtain the relation\begin{align}\la{79}
\langle0|\mathrm{P}_k\rho_{-k}|0\rangle= \frac{\rm \hbar k}{(\ell k)^2}\left(1 - \frac{1}{2}(\frac{1}{2\nu}-1)(k\ell)^2\right
)\langle0|\rho_k\rho_{-k}|0\rangle=\frac 12\hbar {
\rm k} N.
\end{align}
The known result \cite{GMP} for the  spectral factor follows
 \begin{align}\la{80}
s_k=\langle0|\rho_k\rho_{-k}|0\rangle\approx  \frac{1}{2}(k\ell)^2\left(1 + \frac{1}{2}(\frac{1}{2\nu}-1)(k\ell)^2\right)+\dots.
\end{align}
We see that the anomalous  term accounts the universal \({\cal O}(k^4) \)
in the structure factor.

\subsection{Hall conductance in a non-uniform background}\la{D} The formulas of the previous section are readily adapted to study  transport in the topological sector, say a transport in a non-uniform electric field. 

 An electric field acts only on vortices. Thus we add the term \(e\rho E\) in the r.h.s. of the conservation law \eq{611}.\\ 
In the topological sector (\(\Delta_{\nu}\to\infty)\) the flow is steady, and the  only part of the flux tensor survives in the limit is the Lorentz shear force. It balances the Lorentz force\begin{align}
-\nabla _b \sigma_{ab}'=\rho\,( \,e  E - \frac
ec B  \times v)_a
\end{align}   Pulling velocity to the left and in  the leading approximation in gradient  we obtain Eq. \eq{22} of the Sec.
\ref{EE1}.
That equation  yields the universal correction to the Hall conductance \eq{23}.
Comparing the expressions of the spectral function \eq{80} and the Hall conductance we observe a simple relation between the two objects.  It can be obtained from a general theory of the linear response.
% \begin{center}
%\line(1,0){100}
%\end{center}
%This work is a result

%Similar results were obtained by E. Bettelheim by the quantization of the
%Calogero model in the phase space \cite{Eldad}.    
Similar relation  occurs
between the density and a non-uniform magnetic field. A non-uniform magnetic
field enters into the relation   \eq{v3}
 through   the mean density   \(\bar\rho=\frac{\nu}{h
}eB\). At the ground state where velocity vanishes the   \eq{v3}
becomes the Liouville-like equation for the density. In the leading   approximation
in gradients we obtained a generalization of  the Streda formula  \(e\langle
0|\rho_k|0\rangle =\sigma_{xy}(k)B_k  \) for a weakly non-uniform magnetic
field.

\paragraph{Accumulation of charges in curved space.} \smallskip

Anomalous properties of FQHE are   seen in a curved space. Here we mention
just one. In  a curved space the density (the number of particles per unit
area \(\rho\sqrt g dzd\bar z) \) is not uniform but rather depends on the
curvature \begin{align}
\rho=\bar\rho+\frac{1}{4\pi}R+\mathcal{O}(\ell^2\Delta R).\la{261}
\end{align}
 The first term of the gradient expansion in the Gaussian curvature
follows from the shift
formula \eq{v3}
In the curved space the density transformed as \(\rho\to \rho\sqrt g\). Under
this transformation the anomalous term in \(\eq{61}\) acquires an addition
\(\frac{\hbar}{4\nu}\nabla\times\sqrt g\) which yields the term \(-\frac{1}{2\pi}\frac{1}{\sqrt
g}\Delta\log\sqrt g\)  in the
r.h.s. of \eq{61} and subsequently \eq{261}.
Recall that \(R=-\frac{2}{\sqrt
g}\Delta\log\sqrt g\) is the Gaussian curvature. The next term in the expansion
\eq{261} is also universal, but 
requires a more involved analysis.

 Particles/vortices accumulate at curved parts being pushed there by the
Lorentz shear force. For  example, a cone   with the deficit angle \(\alpha\)
possesses  extra \(\alpha/4\pi \) particles located right at the vertex.

Eq.\eq{261} can be checked against the known formula for the number of particles
at the Laughlin state on a Riemannian manifold. Integrating \eq{261} and
using Gauss-Bonnet theorem we obtain   \(N=\nu N_\phi+\frac 12 \chi\) , where
\(\chi\) is Euler characteristic.
\bigskip

Discussions of hydrodynamics of quantum liquids
with I. Rushkin, E. Bettelheim, and T. Can and their
help in understanding the material presented below
are acknowledged. The author thanks the Simons Center for Geometry and Physics for the hospitality during  completion  of the paper. The work was supported by NSF
DMR0820054, DMR 0906427, DMS1206648, BSF2010345 and by John Templeton foundation.


\begin{thebibliography}{10}
\bibitem{C}Our approach is a close analog of the hydrodynamics description
 of Calogero model of Refs. \cite{Calogero}.
\bibitem{mass} In experiments  the cyclotron energy is only  few times larger than 
the gap.
\bibitem{electrostatics}  In  FQH  liquids the Coulomb forces   essentially block propagating waves in the bulk. In this paper we neglect Coulomb 
forces in order to  unmask laws of  quantum hydrodynamics.
\bibitem{FeynmanLandau}See Feynman's
book  \cite{Feynman} p. 319  for an interesting critique  of Landau's
quantum  hydrodynamics approach to the superfluid helium "Proof that superfluidity is not explained by
quantum hydrodynamics".
\end{thebibliography}
\end{document}